\documentclass[journal,twoside,web]{ieeecolor}
\usepackage{etoolbox}
\makeatletter
\@ifundefined{color@begingroup}%
{\let\color@begingroup\relax
\let\color@endgroup\relax}{}%
\def\fix@ieeecolor@hbox#1{%
\hbox{\color@begingroup#1\color@endgroup}}
\patchcmd\@makecaption{\hbox}{\fix@ieeecolor@hbox}{}{\FAILED}
\patchcmd\@makecaption{\hbox}{\fix@ieeecolor@hbox}{}{\FAILED}

\usepackage{tmi}
\usepackage{cite}
\usepackage{amsmath,amssymb,amsfonts}
\usepackage{algorithmic}
\usepackage{graphicx}
\usepackage{textcomp}
\usepackage{graphicx} 
\usepackage{wrapfig} 
\usepackage{placeins} 
\usepackage{multirow}
\usepackage{array}
\usepackage{diagbox}
\usepackage{upgreek}
\def\BibTeX{{\rm B\kern-.05em{\sc i\kern-.025em b}\kern-.08em
    T\kern-.1667em\lower.7ex\hbox{E}\kern-.125emX}}
\begin{document}
\title{Learning a Filtered Backprojection Reconstruction Method for Photoacoustic Computed Tomography with Hemispherical Measurement Geometries}
\author{Panpan Chen, Seonyeong Park, Refik Mert Cam, Hsuan-Kai Huang, Alexander A. Oraevsky, \IEEEmembership{Member, IEEE}, Umberto Villa, \IEEEmembership{Member, IEEE}, and Mark A. Anastasio, \IEEEmembership{Fellow, IEEE}
\thanks{Manuscript received November 28, 2024. This work was supported in part by NIH under Awards R01EB031585 and R01EB034261. (\textit{Corresponding authors: Mark A. Anastasio; Umberto Villa.})}
\thanks{Panpan Chen, Seonyeong Park, and Mark A. Anastasio are with Department of Bioengineering, University of Illinois Urbana–Champaign, Urbana, IL 61801, USA (e-mail: panpanc2@illinois.edu; sp33@illinois.edu; maa@illinois.edu).}
\thanks{Refik Mert Cam and Hsuan-Kai Huang are with Department of Electrical and Computer Engineering, University of Illinois Urbana–Champaign, Urbana, IL 61801, USA (e-mail: rcam2@illinois.edu; hkhuang3@illinois.edu).}
\thanks{Alexander A. Oraevsky is with TomoWave Laboratories, Inc., Houston, TX 77054, USA (e-mail: ao@tomowave.com).}
\thanks{Umberto Villa is with Oden Institute, The University of Texas at Austin, Austin, TX 78712, USA (e-mail: uvilla@austin.utexas.edu).}\\[-4ex]}

\maketitle
\begin{abstract}
In certain three-dimensional (3D) applications of photoacoustic computed tomography (PACT), including \textit{in vivo} breast imaging, hemispherical measurement apertures that enclose the object within their convex hull are employed for data acquisition. Data acquired with such measurement geometries are referred to as \textit{half-scan} data, as only half of a complete spherical measurement aperture is employed. Although previous studies have demonstrated that half-scan data can uniquely and stably reconstruct the sought-after object, no closed-form reconstruction formula for use with half-scan data has been reported. To address this, a semi-analytic reconstruction method in the form of filtered backprojection (FBP), referred to as the half-scan FBP method, is developed in this work. Because the explicit form of the filtering operation in the half-scan FBP method is not currently known, a learning-based method is proposed to approximate it. The proposed method is systematically investigated by use of virtual imaging studies of 3D breast PACT that employ ensembles of numerical breast phantoms and a physics-based model of the data acquisition process. The method is subsequently applied to experimental data acquired in an \textit{in vivo} breast PACT study. The results confirm that the half-scan FBP method can accurately reconstruct 3D images from half-scan data. Importantly, because the sought-after inverse mapping is well-posed, the reconstruction method remains accurate even when applied to data that differ considerably from those employed to learn the filtering operation.
\end{abstract}


% \vspace{3pt}
\section{Introduction}
\label{sec:introduction}
\IEEEPARstart{P}{hotoacoustic} computed tomography (PACT), also known as optoacoustic tomography, is a rapidly developing imaging modality with great potential for a wide range of biomedical applications \cite{wang2012photoacoustic, treeby2010photoacoustic, yao2021perspective}. In PACT, a short laser pulse irradiates the object, leading to the absorption of optical energy \cite{oraevsky2003optoacoustic}. This absorbed energy induces a localized rise in acoustic pressure due to the photoacoustic effect \cite{wang2017photoacoustic}. The resulting acoustic waves travel through the object and surrounding medium and are subsequently measured by ultrasonic transducers. From these measurements, the initial pressure distribution can be estimated by use of a tomographic reconstruction method \cite{li2009photoacoustic, anastasio2007application, poudel2019survey}. PACT offers high optical contrast with inherent functional information in the near-infrared wavelength range, where hemoglobin is one of the major endogenous optical molecular chromophores. It also achieves relatively high resolution, since the photoacoustic signals propagate as acoustic waves, which experience significantly less scattering than light by several orders of magnitude \cite{yao2016multiscale, oraevsky2018full, lin2018single}. Such capabilities enable PACT to provide detailed structural and functional information about the distributions of oxy- and deoxy-hemoglobin concentration in tissues, making it a promising technique for cancer diagnosis and early screening by assessing tumor angiogenesis and hypoxia \cite{weidner1991tumor, lao2008noninvasive, toi2017visualization}. 

A variety of analytic reconstruction methods have been proposed for three-dimensional (3D) PACT \cite{xu2005universal, kunyansky2007series, wang2012simple, finch2004determining, burgholzer2007exact, kunyansky2007explicit, haltmeier2014universal}. Many of these methods employ filtered backprojection (FBP)-type approaches, where reconstruction is facilitated by first filtering measurement data and subsequently backprojecting this filtered data into the image domain \cite{xu2005universal, kunyansky2007explicit, haltmeier2014universal}. The derivations of FBP-based inversion formulae assume that photoacoustic signals are measured over certain canonical closed surfaces that enclose the object \cite{xu2005universal, kunyansky2007explicit, haltmeier2014universal}. In certain modalities, such as 3D PACT breast imaging, physical constraints limit the measurement aperture for data acquisition. For example, in clinical breast imaging applications, hemispherical measurement surfaces \cite{oraevsky2018full} are employed. When standard FBP methods are directly applied to measurement data acquired with hemispherical measurement geometries, the resulting reconstructed images can exhibit inaccuracies and noticeable artifacts\cite{xu2004reconstructions}. Hereafter, the data acquired with such a hemispherical measurement geometry will be referred to as \textit{half-scan} data, with the assumption that the to-be-imaged object resides within the convex hull of the measurement aperture. Similarly, data acquired over a spherical surface that encloses the object will be referred to as \textit{full-scan} data. 

Previous studies have shown that half-scan data uniquely specify the sought-after object \cite{anastasio2005feasibility, ambartsoumian2006range, ambartsoumian2010inversion}. Moreover, results based on microlocal analysis \cite{louis2000local, nguyen2014reconstruction, frikel2015artifacts} have established that half-scan data are sufficient for stable image reconstruction. 
Despite this, direct closed-form methods specifically designed for 3D image reconstruction from half-scan data are currently unavailable. Existing reconstruction methods for half-scan data are either heuristic modifications of direct methods that only partially mitigate artifacts\cite{paltauf2009weight}, or computationally intensive iterative algorithms \cite{pan2003data, anastasio2005feasibility}. Therefore, there remains a need for the development of accurate, direct 3D reconstruction methods for use with half-scan data.

In this work, a semi-analytic FBP-type reconstruction method, referred to as a half-scan FBP method, is developed for accurate and fast image reconstruction from half-scan data. Because the explicit form of the filtering operation in the half-scan FBP method is not currently known, a learning-based method is proposed to approximate it. Key features of the method are that it seeks to approximate a well-posed inverse mapping and involves imaging physics in its formulation. As such, unlike the use of deep learning (DL)-based methods for solving ill-posed problems \cite{antun2020instabilities}, the proposed method is expected to perform reliably even when applied to data that differ considerably from those employed for learning the filtering operation \cite{dehner2023deep, cam2024learning, chen2024learning}. The accuracy of the method is systematically assessed through virtual imaging studies of 3D breast PACT that employ ensembles of numerical breast phantoms and a physics-based model of the data acquisition process. The method is subsequently applied to experimental data from an \textit{in vivo} breast PACT study to further demonstrate its effectiveness under real-world conditions. 

The remainder of the paper is organized as follows. In Section \ref{sec:background}, the PACT imaging model in its continuous and discrete forms is reviewed. The proposed half-scan FBP method is presented in Section \ref{sec:learnedFBP}. The virtual imaging studies performed to validate and investigate the method are described in Section \ref{sec:numerical studies}, with the corresponding results provided in Section \ref{sec:numerical studies results}. In Section \ref{sec:experimental validation}, the method is further validated by use of clinical data from an \emph{in vivo} PACT breast study. Finally, the paper concludes with a discussion in Section \ref{sec:conclusion}.

\section{Background}
\label{sec:background}
In this section, continuous-to-continuous (C-C) and discrete-to-discrete (D-D) imaging models for 3D PACT are reviewed, along with relevant image reconstruction methods.

\vspace{-5pt}
\subsection{Continuous-to-Continuous Imaging Model}
The canonical C-C PACT forward imaging operator maps the sought-after initial pressure distribution $p_0(\textbf{r})$, where $\textbf{r} \in \mathbb{R}^3$, to the acoustic signals \( p(\textbf{r}_0, t)\) that are acquired over a measurement aperture $\Omega_0 \subset \mathbb{R}^3$, where $\textbf{r}_0\in\Omega_0$.
Here, $t \in[0, T]$ is the temporal coordinate confined to the acquisition period $T$, with tissue excitation assumed to occur at $t=0$.
Considering that the to-be-imaged object is acoustically homogeneous and lossless, with acoustic properties matching those of the acoustic coupling medium, 
the canonical forward operator can be described as \cite{wang2013accelerating}:
\begin{equation}
\label{eq:canonical_model}
\scalebox{1.08}{$
p\left(\textbf{r}_0, t\right) =\frac{1}{4 \pi c_0^2} \frac{d}{d t} \int_V d \textbf{r} p_0(\textbf{r}) \frac{\delta\left(t-\frac{|\textbf{r}_0-\textbf{r}|}{c_0}\right)}{|\textbf{r}_0-\textbf{r}|} \\
=\mathcal{H}_{CC} p_0,
$}
\end{equation}
where \( \mathcal{H}_{CC}: \mathbb{L}_2(\mathbb{R}^3) \rightarrow \mathbb{L}_2(\Omega_0 \times [0, T] ) \) denotes the C-C imaging operator, $\delta(t)$ represents the one-dimensional Dirac delta function, $c_0$ is the constant speed of sound (SOS), and $V$ denotes the object support. 

\vspace{-5pt}
\subsection{Discrete-to-Discrete Imaging Model}
\label{subsec:d-d imaging model}
Consider that $N_t$ temporal samples of the pressure signal are recorded with a sampling interval of $\Delta t$ at each of $N_q$ transducer locations on a hemispherical measurement surface that surrounds the to-be-imaged object. Because this work employs an FBP method, it is assumed that the density of transducer locations and temporal sampling frequency are sufficient \cite{haltmeier2016sampling} to avoid significant aliasing artifacts in the reconstructed images. The measured data samples can be lexicographically ordered and represented as the vector $\textbf{p} \in \mathbb{R}^{M \times 1}$, where $M=N_qN_t$. Assuming idealized point-like transducers, the measured signal $[\textbf{p}]_{q N_t+w}$, recorded at time $t = w \Delta t$ by the $q$-th transducer, is expressed as the $(q N_t+w)$-th element of the vector $\textbf{p}$:
\begin{equation}
\label{eq:data-discri}
[\textbf{p}]_{q N_t+w}=\left.p\left(\textbf{r}_0, t\right)\right|_{\textbf{r}_0=\textbf{r}_q, t=w \Delta t},\quad
\begin{array}{c}
\scriptstyle w = 0, \ldots, N_t-1 \\[-0.4em]
\scriptstyle q = 0, \ldots, N_q-1
\end{array}.
\end{equation}

To establish a D-D imaging model, the continuous object $p_0(\textbf{r})$ is approximated by use of a finite-dimensional representation:
\begin{equation}
\label{eq:finite-expansion}
p_0(\textbf{r}) = \sum_{n=0}^{N-1}[\textbf{f}]_n \phi_n(\textbf{r}),
\end{equation}
where $[\textbf{f}]_n$ is the $n$-th element of the coefficient vector $\textbf{f} \in \mathbb{R}^N $, and $\left\{\phi_n(\textbf{r})\right\}_{n=0}^{N-1}$ is a set of given expansion functions. The D-D imaging model that maps the object coefficient vector $\textbf{f}$ to the recorded data samples $\textbf{p}$ can be expressed as \cite{poudel2019survey}:

\begin{equation}
\label{eq:d-d}
\textbf{p}=\textbf{Hf},
\end{equation}
where $\textbf{H}: \mathbb{R}^N \rightarrow \mathbb{R}^M$ is the PACT D-D imaging operator, also referred to as the system matrix. In this study, an interpolation-based D-D imaging model was employed, with the expansion function chosen as piecewise linear basis functions \cite{wang2013accelerating}.
As such, the coefficient vector $\textbf{f}$ can be interpreted as a discretized approximation of the initial pressure distribution. Details regarding the construction of $\mathbf{H}$ are available in \cite{wang2013accelerating}.

\vspace{-5pt}
\subsection{Image Reconstruction from Half-scan Data}
\begin{figure} [ht]
\begin{center}
\begin{tabular}{c} 
\includegraphics[width=0.47\textwidth]{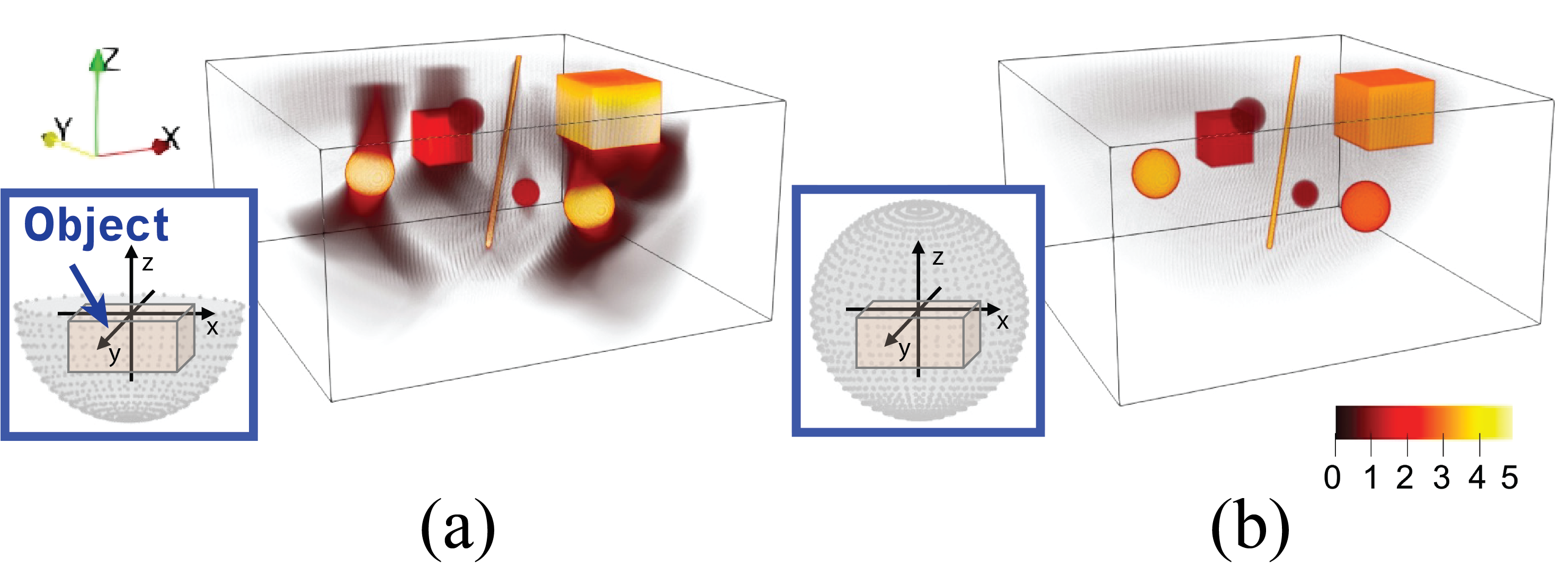}
\end{tabular}
\end{center}
\vspace{-10pt}
\caption[M] 
{ \label{fig:halfUBP} 
Standard FBP \cite{xu2013photoacoustic} results from (a) half-scan data and (b) full-scan data. Insets in (a) and (b) show the measurement geometries used for each data acquisition. In (a), direct application of the standard FBP method to half-scan data results in concentric arc-shaped artifacts originating from the endpoints of the open measurement surface boundary.}
\vspace{-15pt}
\end{figure}

The direct application of analytic methods designed for full-scan data to reconstruct images from half-scan data can result in arc-shaped artifacts and loss of accuracy \cite{frikel2015artifacts}. These artifacts appear as concentric arcs, centered at the endpoints of the open measurement surface and tangent to the object boundary \cite{barannyk2015artifacts}. As illustrated by the example in Fig. \ref{fig:halfUBP} (a), these arc-shaped artifacts are present in the image reconstructed from half-scan data using the standard FBP method, where the object boundaries parallel to the plane formed by the endpoints of the open measurement surface (the x-y plane at z=0) appear blurred.

Despite the fact that closed-form exact reconstruction formulae for use with half-scan data have not been identified, mathematical analyses on the invertibility of the C-C imaging model have yielded encouraging results. Specifically, it has been established that an object can be uniquely \cite{pan2003data, anastasio2005feasibility} and stably \cite{louis2000local} reconstructed from half-scan data. These findings established that the associated inverse problem in an infinite-dimensional vector space setting is well-posed. Iterative reconstruction methods can accurately reconstruct finite-dimensional object estimates from discrete samples of half-scan data \cite{pan2003data, xu2004reconstructions, anastasio2005feasibility}. However, iterative methods can be computationally expensive for 3D problems. As such, there remains an important need for the development of computational efficient and accurate reconstruction methods for use with half-scan data.

\vspace{-5pt}
\section{Learned Half-Scan FBP Method}
\label{sec:learnedFBP}
In this section, a semi-analytic discrete FBP-type method for use with half-scan data is developed. The method, referred to as the \emph{learned half-scan FBP method}, is partially data-driven and designed to ensure accurate and rapid image reconstruction from half-scan data.

\vspace{-7pt}
\subsection{Formulation of a Half-scan Discrete FBP Method}
The discrete form of a half-scan FBP-type reconstruction method can be expressed as
\begin{equation}
\vspace{-2pt}
\label{eq:fbp}
\hat{\textbf{f}} = \textbf{H}^{\dagger} \textbf{F} \textbf{p},
\vspace{-2pt}
\end{equation}
where $\hat{\textbf{f}} \in \mathbb{R}^{N}$ denotes the reconstructed estimate of the object (a discrete approximation of the initial pressure distribution), $\textbf{H}^{\dagger} \in \mathbb{R}^{N \times {M}} $ is the adjoint of the D-D imaging operator, $\textbf{p}$ is the measured half-scan data, and $\textbf{F} \in \mathbb{R}^{{M} \times {M}}$ is the unknown data filtering operator. 
From the singular value decomposition (SVD) of $\textbf{H}$ \cite{cam2024learning}, a mathematical expression for the optimal linear filtering operator $\textbf{F}^{opt} \in \mathbb{R}^{{M} \times {M}}$ can be derived. However, computing the SVD of the large-scale imaging operator $\textbf{H}$ in three dimensions is computationally challenging \cite{saibaba2021randomized}. As a more practical alternative, the optimal filtering operator can be approximated by use of a supervised learning approach \cite{cam2024learning} as described next.

\begin{figure*} [ht]
\begin{center}
\begin{tabular}{c} 
\includegraphics[width=0.93\textwidth]{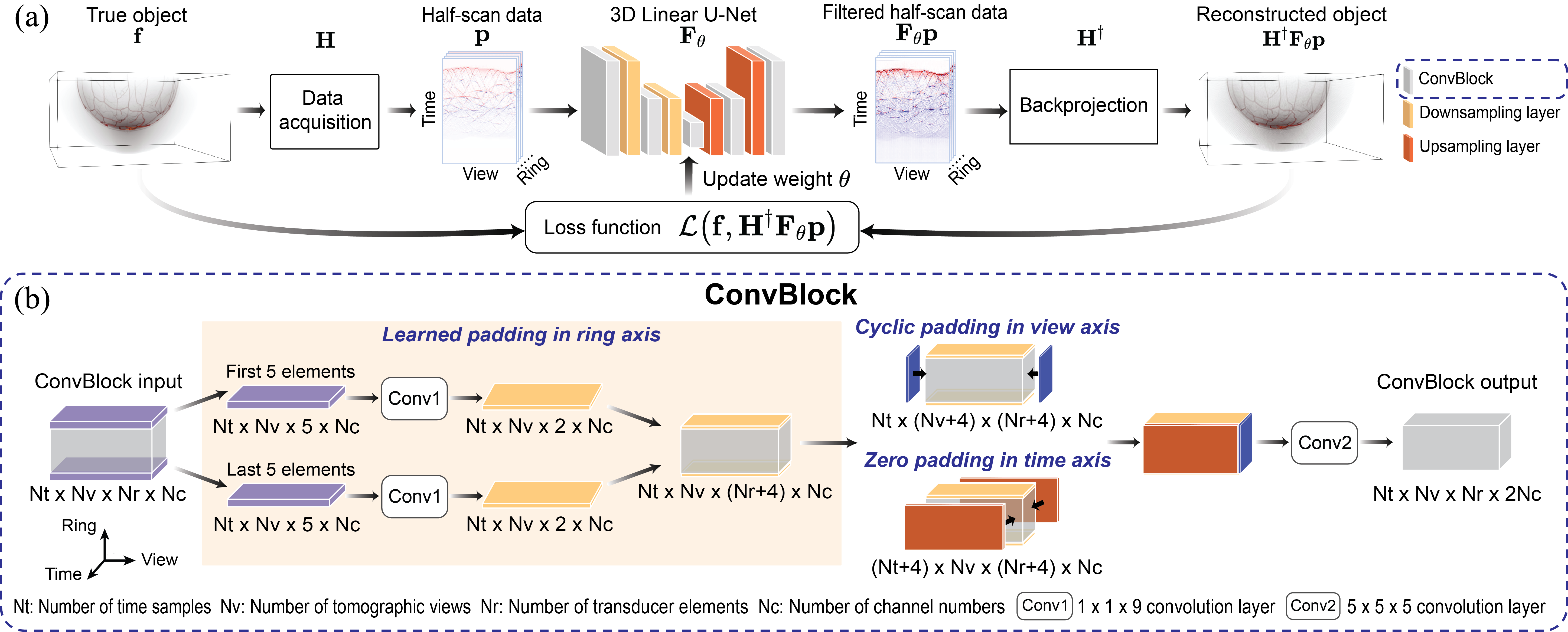}
\end{tabular}
\end{center}
\vspace{-10pt}
\caption[M] 
{ \label{fig:framework}
Learned half-scan FBP framework (a), where a linear 3D U-Net, consisting of convolution blocks (ConvBlocks) as shown in (b) as well as downsampling and upsampling layers, filters the half-scan data. The filtered data are subsequently backprojected to the image space for reconstruction. As illustrated in (b), each ConvBlock incorporates not only a convolution operation (Conv2) for feature extraction but also a padding approach based on prior knowledge of half-scan data, utilizing another convolution operation (Conv1) for learned padding.
}
\vspace{-12pt}
\end{figure*}

\vspace{-7pt}
\subsection{Overview of the Learned Half-scan FBP Method}
In the proposed method, the optimal data filtering operator $\textbf{F}^{opt}$ discussed above is approximated by use of a deep neural network, which is linear with respect to the measured half-scan data $\textbf{p}$. This network is denoted as $\textbf{F}_{\boldsymbol{{\theta}}}: \mathbb{R}^{M} \rightarrow \mathbb{R}^{M}$, where the vector $\boldsymbol{\theta}$ denotes the trainable network parameters. In this case, the task of estimating the optimal filtering operator can be interpreted as determining the parameter vector $\boldsymbol{\theta}$ from a set of training data. Here, the training data consist of arbitrary pairs of objects $\textbf{f}^{(k)}$ and the corresponding half-scan measurement data $\textbf{p}^{(k)}\equiv \textbf{H}\textbf{f}^{(k)}$, where the integer $k$ is an index specifying each training sample. According to \eqref{eq:fbp}, the optimal parameter vector $\hat{\boldsymbol{\theta}}$ should satisfy $\textbf{f}^{(k)}=\textbf{H}^{\dagger} \textbf{F}_{\hat{\boldsymbol{\theta}}} \textbf{p}^{(k)}$, $\forall k$.

Therefore, to train the data filtering network 
$\textbf{F}_{\boldsymbol{\theta}}$, the following optimization problem is approximately solved:
\begin{equation}
\vspace{-2pt}
\label{eq:optimization_problem}
\hat{\boldsymbol{\theta}} = \underset{\boldsymbol{\theta}}{\operatorname{argmin}}\frac{1}{K} \sum_{k=1}^K \mathcal{L}(\textbf{f}^{(k)}, \textbf{H}^{\dagger} 
\textbf{F}_{\boldsymbol{\theta}} \textbf{p}^{(k)}),
\vspace{-2pt}
\end{equation}
where $K$ represents the number of training samples.
The function $\mathcal{L}(\cdot, \cdot)$ is the loss function, defined as the mean squared error (MSE) between the $k$-th reconstruction result $\hat{\textbf{f}}^{(k)} = \textbf{H}^{\dagger}
\textbf{F}_{\boldsymbol{\theta}} \textbf{p}^{(k)}$ and the corresponding true object $\textbf{f}^{(k)}$. 

Once the data filtering network is trained, it can replace $\textbf{F}$ in \eqref{eq:fbp} to form a learned half-scan FBP method, as depicted in Fig. \ref{fig:framework} (a). 
Because of the well-posed nature of the inverse problem and the incorporation of imaging physics through $\textbf{H}^{\dagger}$, the method is expected to perform well even when applied to measurement data that significantly differ from those employed for the filtering network training. In contrast, image-to-image-based DL approaches for image reconstruction are prone to producing image hallucinations\cite{cam2024learning,bhadra2021hallucinations}. These hallucinations can arise due to the heavy reliance of such methods on learning features in the image domain, making them more susceptible to generating false structures.

\vspace{-7pt}
\subsection{Network Architecture in the Learned Half-scan FBP Framework}
\label{sec:architecture}
The following details regarding the assumed imaging system configuration are relevant to the specification of the data filtering network. Motivated by an existing 3D PACT breast imager \cite{oraevsky2018full}, the system was assumed to comprise an arc-shaped acoustic measurement probe that was rotated in discrete steps to acquire data samples on a hemispherical measurement aperture. This configuration is parameterized by $N_r$, $N_v$, and $N_t$, which denote the number of transducer elements on the arc-shaped probe (ring dimension), the number of tomographic views (view dimension), and the number of time samples (time dimension), respectively. 

\subsubsection{Linear 3D U-Net Architecture}
Given that the half-scan data $\textbf{p} \in \mathbb{R}^{N_tN_q\times 1}$, with $N_q=N_vN_r$, can be represented as a 3D tensor, a 3D U-Net model was employed to approximate the existing but unknown filtering operator. The 3D U-Net model was chosen as a canonical example of a deep neural network that has been widely deployed in medical imaging applications \cite{kofler2018u, chen2020improved}; however, alternative network models can also be employed for data filtering. Moreover, for the considered problem, the sought-after inverse mapping is inherently linear. Therefore, a linear version of the 3D U-Net was adopted by excluding all nonlinear structures.

The employed linear 3D U-Net consisted of encoder and decoder modules. The encoder was comprised of a series of convolution blocks (ConvBlocks) and downsampling layers, enabling the extraction of multi-scale feature representations. The downsampling layer was implemented using a 3D convolution with a kernel size of 3$\times$3$\times$3, a stride of 2, and an output channel size equal to the input channel size. As illustrated in Fig. \ref{fig:framework} (b), each ConvBlock consisted of two convolution operations: one for learned padding in the ring dimension (Conv1), which will be detailed later, and another for feature extraction with a 5$\times$5$\times$5 kernel (Conv2). The first ConvBlock contained 8 feature channels, doubling with each subsequent ConvBlock, leading to 256 channels at the bottleneck ConvBlock at the end of the encoder. The decoder module, following the bottleneck, consisted of ConvBlocks and upsampling layers. Transposed convolution was employed for upsampling, with a kernel size of 2$\times$2$\times$2, a stride of 2, and the same number of output channels as the input channels. The number of feature channels was halved after each upsampling step, and the resulting map was concatenated with the corresponding feature map from the encoder at the same level. Finally, a 1$\times$1$\times$1 convolution was applied to produce the single-channel U-Net output.

\vspace{-5pt}
\subsection{Padding Approach for Half-scan Data}
To preserve the input dimensions through ConvBlocks, padding must be applied at the boundaries of the input data. However, the commonly used zero padding can introduce inconsistencies near these boundaries, potentially affecting the learning process for the data filtering network. To maintain consistent features of the input, the padding strategy in ConvBlocks was designed based on the configuration of the assumed imaging system. This approach effectively approximates data that are physically unmeasured at the padding positions, as illustrated in Fig. \ref{fig:framework} (b). For the view dimension, where tomographic views exhibited 2$\pi$ periodicity, cyclic padding was employed to maintain azimuthal continuity. In the time dimension, assuming temporally untruncated measurement, zero padding was applied, as pressure traces outside the measured time range were considered negligible. Along the ring dimension, a learned padding approach was implemented, employing learned values to account for the unmeasured data beyond the transducer arc. This process began by extracting the first and last five elements of the input data along the ring axis, which were then processed through 1$\times$1$\times$9 convolution layers (Conv1). The outputs from Conv1 were subsequently concatenated with the original input data along the ring axis. 

\section{Virtual Imaging Studies}
\label{sec:numerical studies}
This section outlines the virtual imaging studies conducted to validate the learned half-scan FBP method. These studies assess the method's accuracy on a dataset whose characteristics match the training data, hereafter referred to as the in-distribution (ID) test set, and its generalizability on datasets that differ from the training data, referred to as out-of-distribution (OOD) test sets. To generate these datasets, two types of to-be-imaged objects were utilized: stochastically generated anatomically realistic 3D numerical breast phantoms (NBPs) \cite{park2023stochastic} and 3D mouse phantoms \cite{segars2009mcat}, both incorporating realistic functional and optical properties. Six datasets were produced, varying by imaging system configurations and noise considerations. One of them was employed as the ID dataset, and the remaining five were used as OOD test sets. Further details are provided below.

\vspace{-7pt}
\subsection{Virtual PACT Imaging and Data Acquisition}
\subsubsection{Virtual Imaging System Configuration}
\label{sec:vis_config}
Three different virtual 3D PACT imaging systems were configured to emulate real-world imaging scenarios, each differing in their optical illumination subsystem design. The first system, referred to as \textit{System A}, employed 20 evenly positioned arc-shaped illuminators (radius of 145 mm, central angle of 80$^\circ$), each equipped with five linear fiber-optic segments along the arc-shaped surface. Each segment was represented by a broadening slit light source (half-angle of 12.5$^\circ$). The second system, \textit{System B}, used a single cone beam (half angle of 36$^\circ$) to illuminate the object from the bottom upwards, while the third system, \textit{System C}, applied uniform illumination across the object surface.
All three systems employed an identical acoustic measurement subsystem. Acoustic signals were detected by 107 idealized point-like transducer elements uniformly distributed along the surface of an arc-shaped photoacoustic probe (radius of 85 mm, central angle of 90$^\circ$), with an angle of approximately 0.84$^\circ$ between adjacent elements. The probe was rotated to form a complete hemispherical aperture, collecting data across 320 tomographic views during scanning. The number of measured time samples was set to 1280, with a sampling frequency of 10 MHz. 

\subsubsection{Simulation of the Photoacoustic Effect}
To establish the 3D initial pressure distributions representing the to-be-imaged objects in the acoustic simulation, the optical energy deposition in the numerical phantoms was simulated using the GPU-accelerated MCX software \cite{fang2009monte} for Systems A and B, and the diffusion approximation \cite{tarvainen2012reconstructing} for System C. In the virtual imaging Systems A and B, each fiber-optic segment and the single cone beam were simulated using $10^{10}$ and $5\times10^{10}$ photons, respectively, both over 5 ns. An illumination wavelength was randomly selected from 757 nm, 800 nm, and 850 nm for each object in both systems. Virtual imaging System C involved uniform illumination on the object surface at a wavelength of 800 nm.

\vspace{1pt}
\subsubsection{Simulation of Acoustic Measurement Acquisition}
Acoustic measurements were simulated by use of the interpolation-model-based D-D forward model described by \eqref{eq:d-d}. A reference SOS of 1509.15 m/s was assumed, which was chosen based on experimental breast data used in the \textit{in vivo} study presented in Section \ref{sec:experimental validation}. Full-scan data were also simulated for use in reconstructing high-quality reference images. In the full-scan data acquisition, which is not physically realizable with \textit{in vivo} breast PACT, a spherical aperture was formed by scanning with an arc-shaped probe (radius of 85 mm, central angle of 180$^\circ$) equipped with 213 evenly distributed transducer elements, each spaced approximately 0.84$^\circ$ apart, and rotated in 320 discrete steps. Additionally, noisy versions of the measurements were produced. Additive noise was modeled as an independent and identically distributed Gaussian random variable with a mean of zero and a standard deviation equal to 1\% of the maximum photoacoustic signal amplitude observed across all training, validation, and ID testing sets \cite{park2022normalization}. These datasets are detailed in the next subsection.

\begin{figure} [ht]
\begin{center}
\begin{tabular}{c} 
\includegraphics[width=0.47\textwidth]{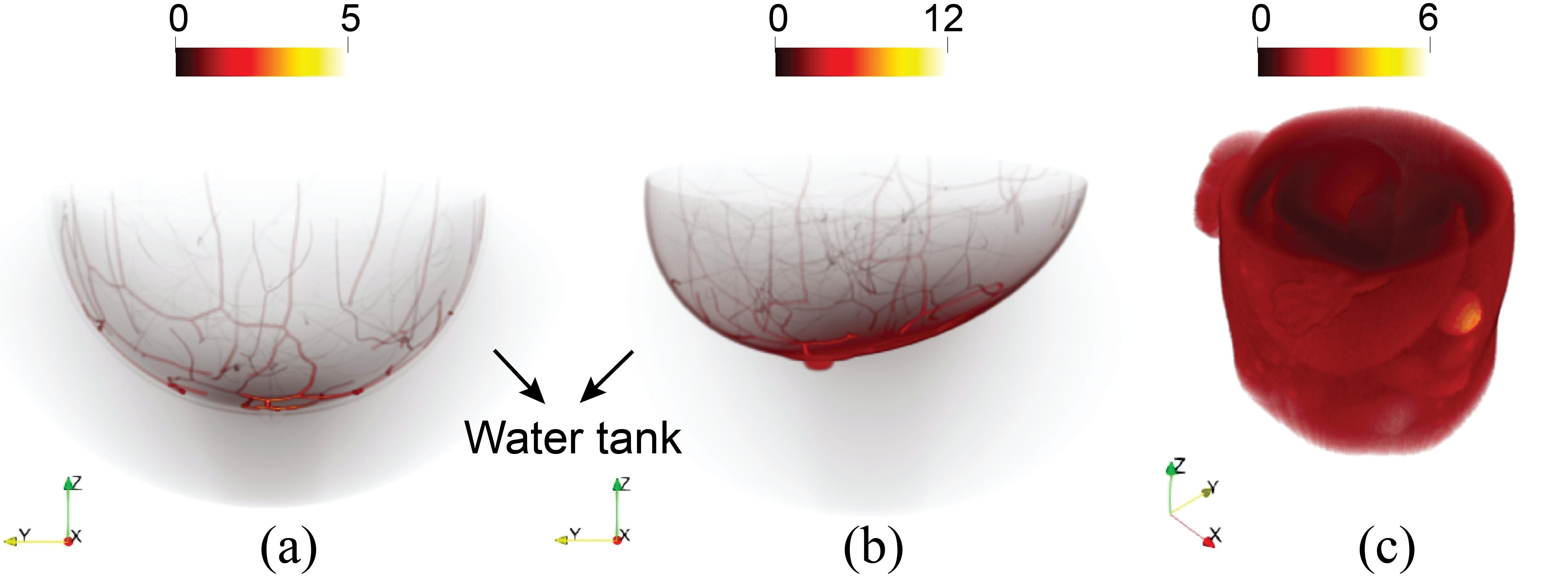}
\end{tabular}
\end{center}
\vspace{-14pt}
\caption[M] 
{ \label{fig:dataset} 
Examples of 3D initial pressure distributions from (a) NBP-A (b) NBP-B, and (c) MOBY datasets. Volume rendering was performed using Paraview \cite{ahrens200536}, which accumulates intensities based on the chosen color and opacity maps.}
\vspace{-5pt}
\end{figure}

\subsubsection{Training, Validation, and Test Datasets}
Six distinct ensembles were simulated, pairing three types of initial pressure distributions with corresponding PACT measurements under either noiseless or noisy conditions. Figure \ref{fig:dataset} illustrates examples of the three types of initial pressure distributions. The ensembles, derived from NBPs using the virtual imaging System A setup, are referred to as the \textit{noiseless NBP-A} and \textit{noisy NBP-A} datasets. Similarly, those generated from NBPs with the virtual imaging System B setup constitute the \textit{noiseless NBP-B} and \textit{noisy NBP-B} datasets. The final ensembles, simulated using abdomen-region-cropped 3D numerical mouse whole-body phantoms (MOBYs) \cite{segars2009mcat} with the virtual imaging System C setup, are referred to as the \textit{noiseless MOBY} and \textit{noisy MOBY} datasets. The learned half-scan FBP method was trained on the noiseless NBP-A dataset. To evaluate the method's generalizability, five OOD test sets were employed. One of these was the noisy NBP-A dataset, which utilized the same type of to-be-imaged objects as the training data but under noisy conditions. The remaining sets were the noiseless and noisy NBP-B and MOBY datasets, which represent more challenging scenarios as their to-be-imaged objects differed significantly from those used for training. A total of 5000 noiseless NBP-A samples were divided into 4500 training, 250 validation, and 250 test samples. The OOD test sets consisted of 250 noisy NBP-A, 80 noiseless and noisy NBP-B, and 10 noiseless and noisy MOBY samples.

\begin{figure*} [ht]
\begin{center}
\begin{tabular}{c} 
\includegraphics[width=0.93\textwidth]{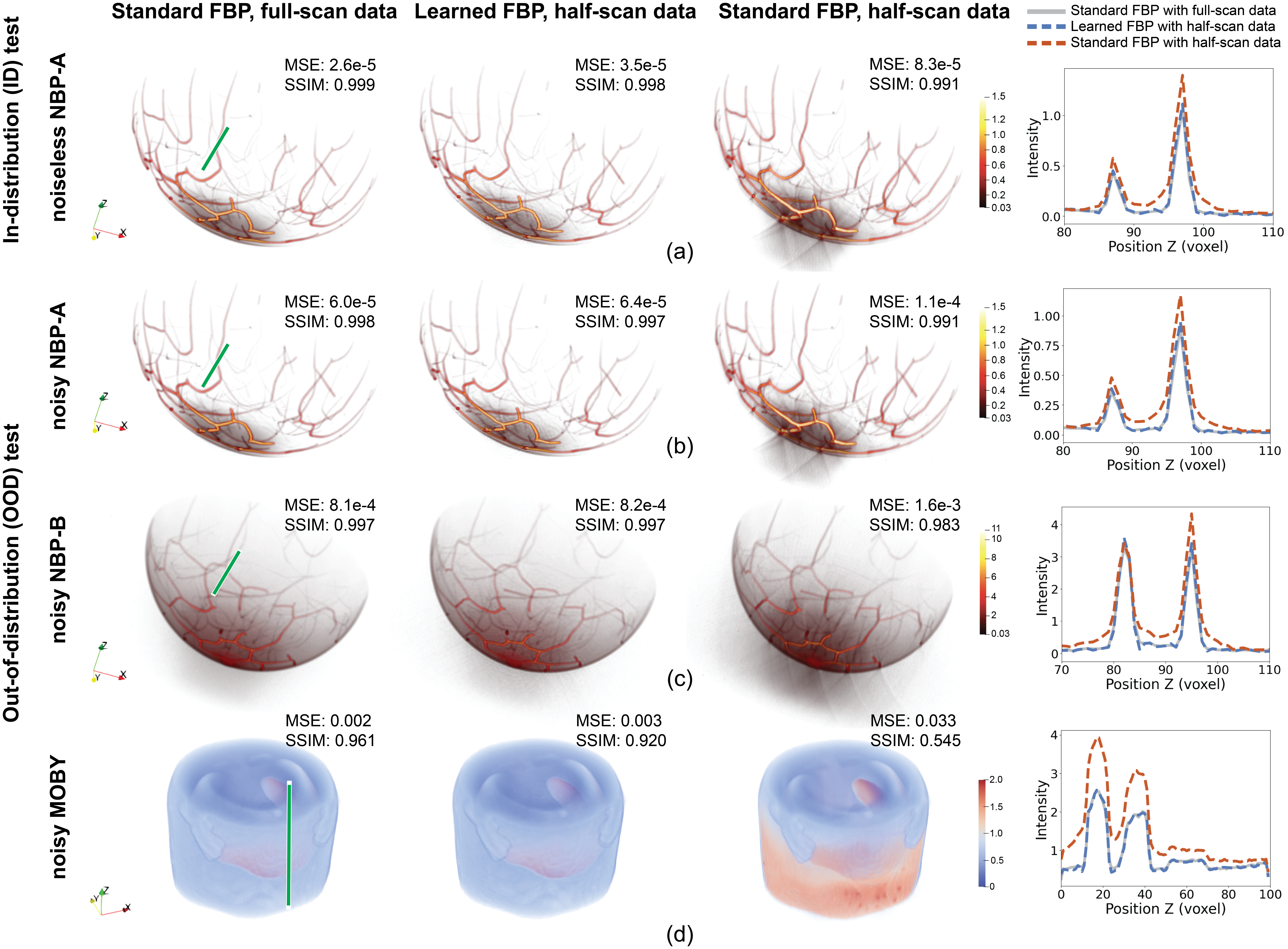}
\end{tabular}
\end{center}
\vspace{-10pt}
\caption[M] 
{\label{fig:visulization}{Visual inspections (first to third columns) and line profile comparisons (fourth column) for the ID test from (a) the noiseless NBP-A dataset and for the OOD tests from (b) the noisy NBP-A, (c) noisy NBP-B, and (d) noisy MOBY datasets. The first to third columns display the volumes reconstructed using the standard FBP method applied to full-scan data, the learned half-scan FBP method, and the standard FBP method applied to half-scan data, respectively. Volume rendering was performed using Paraview. The line along which the profiles compared in the rightmost column were extracted is annotated in the volumes contained in the first column, indicated by green lines. In the profile comparisons (fourth column), the gray solid line, blue dashed line, and red dashed lines correspond to the standard FBP method applied to full-scan data, the learned half-scan FBP method, and the standard FBP method applied to half-scan data, respectively. The comparisons reveal that the learned half-scan FBP method significantly outperformed the standard FBP method applied to half-scan data and performed comparably to the standard FBP method applied to full-scan data, for both ID and OOD test samples}, even in the presence of noise.}
\vspace{-12pt}
\end{figure*}

\vspace{-7pt}
\subsection{Implementation of the Learned Half-scan FBP}
The implementation of the learned half-scan FBP method involves a 3D linear U-Net as the data filtering model and the backprojection operation $\textbf{H}^\dagger \in \mathbb{R}^{N \times M}$. Here, $M$ was set to 1280 $\times$ 320 $\times$ 107, based on the total number of time samples, tomographic views, and transducer elements, as described in Section \ref{sec:vis_config}. The number of elements representing the discretized object $N$ was set to 340 $\times$ 340 $\times$ 170, corresponding to a physical volume of 170 $\times$ 170 $\times$ 85 $\text{mm}^3$ with a voxel size of 0.5 mm. This voxel size, along with the sampling frequency of 10 MHz in the acoustic measurement system, was chosen based on the available memory in the employed GPU. The interpolation-model-based D-D forward operation $\textbf{H}$ and backprojection operation $\textbf{H}^\dagger$ were implemented in C++ and accelerated using the CUDA library \cite{wang2013accelerating}.

The 3D linear U-Net-based data filtering model $\textbf{F}_{\boldsymbol{\theta}}$ was implemented in PyTorch \cite{paszke2017automatic}. To enable automatic differentiation of the loss function in \eqref{eq:optimization_problem}, a user-defined PyTorch class was implemented to evaluate the actions of $\textbf{H}$ and $\textbf{H}^\dagger$. The Adam optimizer was utilized for training, with a learning rate set to $10^{-4}$ and a batch size of 1. To accelerate training, the loss function was calculated from partial object volumes sized 340 $\times$ 85 $\times$ 170 voxels, randomly selected along the y-axis.

The model was trained on an eight-way A100 GPU compute node in the Delta advanced computing resource \cite{gropp2023delta}, utilizing 8 NVIDIA A100 GPUs and an AMD 64-core 2.55 GHz Milan processor. The testing of the proposed method, as well as the implementation of other reference methods, was conducted on a system equipped with 4 NVIDIA A100 GPUs and an Intel\textsuperscript{\textregistered} Xeon\textsuperscript{\textregistered} Silver 4309Y CPU @ 2.80GHz. 

\vspace{-7pt}
\subsection{Study Designs and Evaluations}
The following studies were performed to assess the accuracy, stability, and generalizability of the learned half-scan FBP method. First, the method was established by training the data filtering network on the noiseless NBP-A training set, and its reconstruction performance was evaluated on the noiseless NBP-A test set. Without retraining, the method was then tested on five OOD test sets: the noisy NBP-A dataset, as well as the noiseless and noisy NBP-B and MOBY datasets. The reconstruction accuracy of the learned half-scan FBP method was compared to that of the standard FBP method applied to both half-scan data and full-scan data. In all cases involving noisy data, a Gaussian low-pass filter with a half width half maximum (HWHM) of 0.1177$\mu$s was implemented in SciPy \cite{2020SciPy-NMeth} and applied to the simulated noisy measured data prior to image reconstruction. This is equivalent to apodizing the data filtering operation, which is a common approach to mitigating noise amplification in FBP-type methods \cite{xu2005universal}.

The learned half-scan FBP method was qualitatively and quantitatively assessed. Visual inspections were conducted through volume rendering and line profile comparisons. Reconstructed images were quantitatively assessed by use of MSE and the structural similarity index (SSIM), which were calculated with respect to the true objects. These assessments were performed using both ID and OOD testing datasets.

\vspace{-5pt}
\section{Virtual Imaging Results}
\label{sec:numerical studies results}
\begin{figure*} [ht]
\begin{center}
\begin{tabular}{c} 
\includegraphics[width=0.97\textwidth]{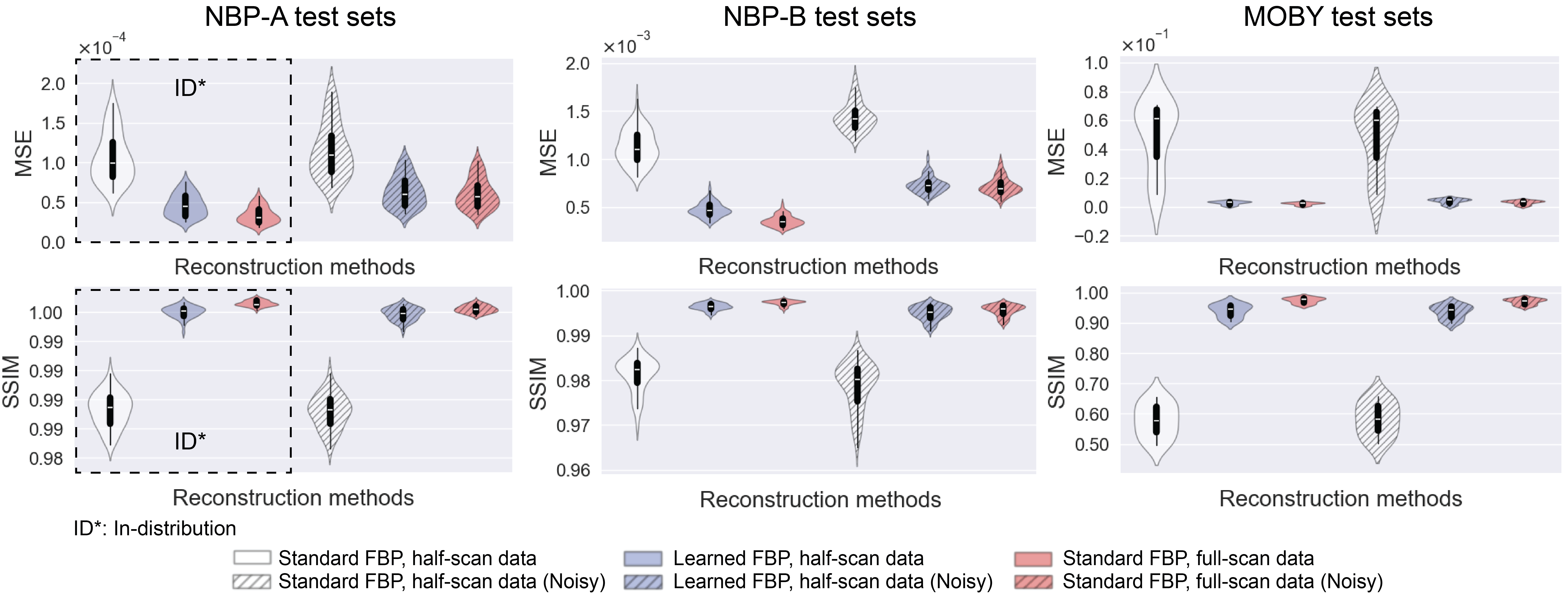}
\end{tabular}
\end{center}
\vspace{-10pt}
\caption[M] 
{ \label{fig:in-distribution-violin} 
Violin plots of the MSE (top row) and SSIM (bottom row) values for the standard FBP method applied to half-scan data (white), the learned half-scan FBP method (blue), and the standard FBP method applied to full-scan data (red) under noiseless (solid fill) and noisy conditions (diagonally striped fill). Results are shown for 250 NBP-A (first column), 80 NBP-B (second column), and 10 MOBY samples (third column) for each method. The dashed box in the first column indicates the results from the ID test, while the remaining plots represent the OOD test results. The quantitative results indicate that the learned half-scan FBP method (blue) is robust in noisy conditions (diagonally striped fill) and exhibits strong generalizability across different types of objects (second and third columns), achieving a reconstruction accuracy comparable to the standard FBP method applied to full-scan data (red).}
\vspace{-12pt}
\end{figure*}

Figures \ref{fig:visulization} (a) and (b) present examples of images reconstructed using the noiseless and noisy NBP-A data, which correspond to the ID and OOD test results, respectively. Both visual inspection and line profiles confirmed that the images reconstructed by use of the standard FBP method applied to full-scan data (left column) and the learned half-scan FBP method (second column) were nearly indistinguishable, which was consistent with the similar MSE and SSIM values reported in the figures. This confirmed that the proposed method can reconstruct images from half-scan data that closely approximate ones reconstructed using the standard FBP method applied to full-scan data, even in the presence of measurement noise. On the other hand, both the noiseless and noisy images reconstructed by use of the standard FBP method from half-scan data (third column) exhibited arc-shaped artifacts, consistent with the predictions of the microlocal analysis findings mentioned in Section \ref{subsec:d-d imaging model}. These artifacts caused a widening of the object structures along the z-axis, as also revealed by the provided line profile comparisons. Additionally, the MSE and SSIM values for these images were found to be inferior to those corresponding to the learned half-scan FBP reconstructions. 

Figures \ref{fig:visulization} (c) and (d) illustrate the OOD test results from noisy measurements simulated using the NBP-B and MOBY datasets, respectively. As in the ID testing results, the learned half-scan FBP method (second column) performed comparably to the standard FBP method applied to full-scan data (first column) and outperformed the standard FBP method applied to half-scan data (third column), substantially reducing artifacts. These conclusions are further corroborated by the line profile comparisons and the reported MSE and SSIM values.

Figure \ref{fig:in-distribution-violin} presents violin plots of the MSE and SSIM values for the standard FBP method applied to half-scan data (white), the learned half-scan FBP method (blue), and the standard FBP method applied to full-scan data (red), under noiseless (solid fill) and noisy conditions (diagonally striped fill). Results are shown for 250 NBP-A, 80 NBP-B, and 10 MOBY samples for each method. Consistent with the results in Fig. \ref{fig:visulization}, the proposed method demonstrated robustness in noisy conditions and strong generalizability across different types of objects.

\vspace{-5pt}
\section{Application to Experimental Breast Data}
\label{sec:experimental validation}
The performance of the learned half-scan FBP method under real-world conditions was explored by use of experimental breast PACT data acquired in an \textit{in vivo} study.

\vspace{-7pt}
\subsection{Data Acquisition}
In this study, archived, fully-anonymized experimental breast data \cite{oraevsky2018full} were utilized. These data were acquired from both the left and right breasts of a healthy female volunteer at an illumination wavelength of 755 nm. Data collection was performed using the LOUISA-3D system (TomoWave Laboratories, Houston, TX) \cite{oraevsky2018full} at MD Anderson Cancer Center. The LOUISA-3D system, whose acoustic detection system was modeled in the virtual imaging studies, was equipped with a photoacoustic probe (radius of 85 mm, center angle of 80$^\circ$) containing 96 ultra-wideband (50 kHz to 6 MHz) transducer elements, each with a square shape sized $1.1 \times 1.1$ mm$^2$ and spaced approximately 0.84$^\circ$ apart. By rotating the probe in discrete steps, a spherical segment aperture was formed. In this experimental study, 1536 pressure time samples were collected over 76.8$\mu$s at a sampling frequency of 20 MHz \cite{oraevsky2018full}.

\begin{figure*} [h]
\begin{center}
\begin{tabular}{c} 
\includegraphics[width=0.93\textwidth]{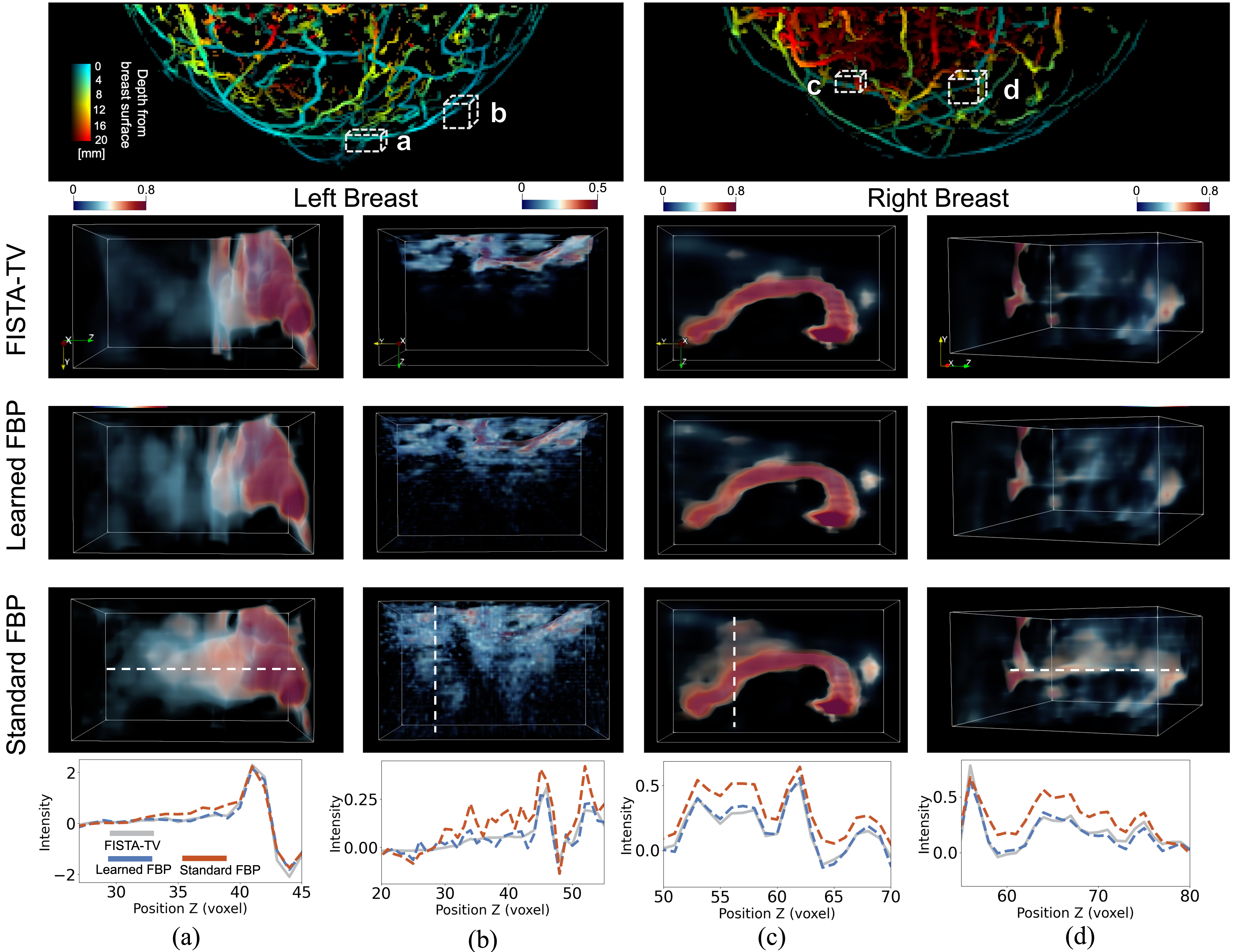 }
\end{tabular}
\end{center}
\vspace{-12pt}
\caption[M] 
{\label{fig:exp-combine} 
Visual inspection of images reconstructed from \textit{in vivo} breast data using three methods: the FISTA-TV, learned half-scan FBP, and standard FBP methods. The top row, from left to right, displays images of the entire left (a and b) and right breasts (c and d), reconstructed using the learned half-scan FBP method with optical fluence normalization \cite{park2022normalization} and depth-based color coding applied. The second to fourth rows show zoomed-in regions of the images reconstructed by use of each method, respectively. The selected regions are highlighted by white boxes in the top row. White dashed lines in the zoomed-in regions in the fourth row indicate the locations of profiles compared in the bottom row. The results demonstrate that the learned half-scan FBP method generalizes effectively to experimental data, mitigating artifacts and inaccuracies observed when the standard FBP method is applied to half-scan data.
}
\vspace{-12pt}
\end{figure*}

\vspace{-10pt}
\subsection{Image Reconstruction and Evaluation}
The learned half-scan FBP method employing a data filter $\textbf{F}_{\boldsymbol{\theta}}$ that was trained on the noiseless NBP-A dataset in the virtual imaging studies, was applied to \emph{in vivo} data. To accommodate for differences in the data acquisition design between the LOUISA-3D system and the virtual imaging system used to train the filter, the following preprocessing steps were applied for the learned half-scan FBP method. First, a Gaussian low-pass filter with a HWHM of 0.05887$\mu$s was applied to the experimental data to mitigate noise. Next, experimental data were subsampled at a sampling frequency of 10 MHz to match that used in the virtual imager. Finally, experimental data were zero-padded along the time and transducer ring dimensions to match the number of time samples and transducer rings expected by the pretrained filter $\textbf{F}_{\boldsymbol{\theta}}$. 

To evaluate the performance of the learned half-scan FBP method, two reference methods were considered: the standard FBP method applied to the acquired half-scan data and an iterative method in which a penalized least squares problem with total variation regularization was solved using the Fast Iterative Shrinkage-Thresholding Algorithm, referred to as \textit{FISTA-TV} \cite{beck2009fast}. 
For the standard FBP method, the following preprocessing steps were applied. The experimental data were first smoothed with a Gaussian low-pass filter with a HWHM of 0.05887 $\mu$s and then subsampled to a sampling frequency of 10 MHz. For the FISTA-TV method, no preprocessing was applied to the experimental data. The regularization parameter was set to $4 \times 10^{-3}$ after a parameter sweep within the range [$1\times 10^{-4}$, $1\times 10^{-2}$] and chosen based on visual examination, with a balance between noise suppression and the preservation of image details as the selection criterion. The algorithm terminated when $||\textbf{f}^{(i)} – \textbf{f}^{(i - 1)}||^2/\max_{l \leq i} ||\textbf{f}^{(l)} – \textbf{f}^{(l - 1)}||^2$ fell below a threshold of $5 \times 10^{-2}$, where $i$ and $l$ denote the iteration index \cite{cam2024spatiotemporal}. For image reconstruction of the left and right breasts, constant SOS values of 1509.15 m/s and 1503.00 m/s were subjectively selected, respectively.

\vspace{-7pt}
\subsection{Experimental Results}
Figure \ref{fig:exp-combine} presents images that were reconstructed the from experimental data. The top row of Fig. \ref{fig:exp-combine} shows images of the entire left and right breasts, from left to right, reconstructed using the learned half-scan FBP method. For display purposes, optical fluence normalization and depth-based color coding were applied \cite{park2022normalization}. These images also indicate the locations of zoomed-in regions for method comparisons ((a) and (b) for the left breast, (c) and (d) for the right breast), marked with dashed boxes. The second, third, and fourth rows display the zoomed-in regions reconstructed from half-scan data using the FISTA-TV method, the learned half-scan FBP method, and the standard FBP method, respectively. The bottom row shows line profiles from the zoom-in regions, with white dashed lines in the fourth row indicating their locations.
Compared to the reference images reconstructed by use of the FISTA-TV method, the image reconstructed using the standard FBP method from half-scan data exhibited artifacts and false connectivity in the object structure. In contrast, the proposed method effectively suppressed these artifacts, achieving relatively high reconstruction accuracy. The line profile comparison in the bottom row of Fig. \ref{fig:exp-combine} further substantiates these findings.

Table \ref{tab: exp-metrics} provides the MSE and SSIM values of images reconstructed by use of the learned half-scan and standard FBP methods, relative to the images reconstructed using FISTA-TV that served as references. The average reconstruction times are also provided. The learned half-scan FBP method achieved performance comparable to the FISTA-TV method, while reducing reconstruction times by a factor of 1000. On the other hand, the standard FBP method yielded inferior MSE and SSIM values compared to the learned half-scan FBP method. It is noteworthy that out of the 30 seconds required by the learned half-scan FBP method, only 2 seconds were dedicated to the filtering operation. The remaining time was spent on the interpolation-model-based backprojection operation, which can vary depending on the implementation.
\begin{table}[ht]
\vspace{-7pt}
    \centering
    \caption{MSE, SSIM, and Average Reconstruction Time}
    \label{tab: exp-metrics} 
    \resizebox{\columnwidth}{!}{ 
    \begin{tabular}{p{0.35\columnwidth}|c|c|c}
        \hline
        Method & MSE & SSIM & Average \\
         & (Left, Right breast) & (Left, Right breast) & time \\ 
        \hline  \hline
        Standard FBP & 5.3e-3, 1.1e-3 & 0.977, 0.973 & 3 s \\
        \hline
        Learned half-scan FBP & 4.7e-3, 9.4e-4 & 0.986, 0.983 & 30 s\\ 
        \hline
        Backprojection model & \multirow{2}{*}{1.1e-2, 2.5e-3} & \multirow{2}{*}{0.953, 0.953} & \multirow{2}{*}{28 s} \\
        of Learned half-scan FBP &  &  & \\ 
        \hline
    \multicolumn{4}{p{\columnwidth}}{The FISTA-TV reconstruction took approximately 8 hours for each breast.} \\
    \end{tabular}
    }
\vspace{-7pt}
\end{table}

\vspace{-5pt}
\section{Discussion and Conclusion}
\label{sec:conclusion}
Prior studies have established that PACT image reconstruction from measured data acquired with a hemispherical measurement geometry, i.e., half-scan data, corresponds to a well-posed problem. However, no closed-form inversion formula for this problem is currently available. To address this, the learned half-scan FBP method was proposed, in which a linear deep neural network approximates the unknown data filtering operation for half-scan data.

The learned half-scan FBP method achieved high accuracy, even when applied to test data that differed substantially from the training data, as demonstrated by the numerical and experimental studies. Specifically, it was demonstrated that the proposed method achieved reconstruction accuracy comparable to the FISTA-TV reference method, while accelerating image reconstruction by a factor of 1000. More importantly, the learned half-scan FBP method exhibited robust generalizability, even when applied to experimental breast data acquired in an \textit{in vivo} study.

The robust generalizability of the proposed method stems from the fact that inverse problem associated with PACT image reconstruction from half-scan data is well-posed. A recent work by Cam et al. \cite{cam2024learning} employed a similar strategy, in which a learned FBP method was established for approximately inverting the so-called half-time circular Radon transform. Perhaps the first study that proposed a learned FBP method for a well-posed problem was reported by Floyd \cite{floyd1991artificial}. In such applications, the instabilities commonly observed in DL-based image reconstruction approaches\cite{gottschling2020troublesome} are avoided, resulting in reliable and trustworth methods. In addition, learned FBP methods \cite{wurfl2018deep, shen2024physics} have been investigated, which implement both filtering and backprojection operations via neural networks. 

The proposed method demonstrated reconstruction accuracy comparable to the reference methods. However, in the virtual imaging studies where full-scan data were available, small discrepancies in MSE and SSIM values were observed between the learned half-scan FBP method and the standard FBP method applied with full-scan data, as shown in Fig. \ref{fig:in-distribution-violin}. These discrepancies may have resulted from the fact that, having a learned filtering component, the proposed method only approximates the true inverse mapping. Furthermore, full-scan data contain redundancies, whereas half-scan data do not \cite{anastasio2005feasibility}, which will result in numerical errors and noise being propagated differently into image space.

The half-scan data for network training were simulated with a reference SOS of 1509.15 m/s, chosen empirically. To confirm the trained network's robustness across a range of SOS values possible in \textit{in vivo} breast imaging, the MSE and SSIM values for the learned half-scan FBP method were evaluated within the reported range of breast tissue SOS values (1447 m/s to 1555 m/s) \cite{hopp2012breast}. The resulting MSE values were $4.6\times 10^{-5} \pm 1.6\times 10^{-6}$ and SSIM values were $0.997 \pm 2\times 10^{-4}$, demonstrating the method's robustness across this SOS range. Therefore, for soft tissue PACT imaging where a homogeneous SOS assumption is generally useful, the DL model can be trained with a reference SOS and is expected to maintain high reconstruction accuracy across the SOS range without the need to retrain the model.

Although this study was conducted with a specific PACT imager design, the proposed method can be readily adapted to other PACT imaging system designs. In addition, other D-D imaging operators and their implementations can be adopted, and any suitable DL model can be used for half-scan data filtering. 

\vspace{-7pt}
\section*{Acknowledgement}
This research used the Delta advanced computing and data resource which is supported by the National Science Foundation (award OAC 2005572) and the State of Illinois. Delta is a joint effort of the University of Illinois Urbana-Champaign and its National Center for Supercomputing Applications.

\vspace{-7pt}
\bibliographystyle{IEEEtran}
\bibliography{reference.bib}

\end{document}